\documentclass{ws-procs9x6}

\begin{document}

\title{CANONICAL TENSOR MODEL WITH LOCAL TIME \\AND ITS UNIQUENESS}

\author{N. SASAKURA}

\address{Yukawa Institute for Theoretical Physics, Kyoto University,\\
Kyoto, 606-8502, Japan\\
E-mail: sasakura@yukawa.kyoto-u.ac.jp}

\begin{abstract}
A canonical formalism of the rank-three tensor model with the notion of local time is proposed.
The consistency of the local time evolution is guaranteed by imposing that local Hamiltonians 
and the $so(N)$ kinematical symmetry of the tensor model should form a first class constraint algebra. 
By imposing some physically reasonable assumptions, it is shown that there exist only two 
such local 
Hamiltonians with a slight difference in index contraction. The first class constraint
algebra is shown to approach the DeWitt constraint algebra of the general relativity in a certain
locality limit. Quantization of the system is briefly discussed.
\end{abstract}


\bodymatter

\section{Introduction}\label{sec:introduction}
Formulation of quantum gravity is remaining as one of the major difficulties in physics. 
An interesting direction is to pursue the possibility that the space-time
 is an emergent phenomenon.
The idea of emergent space-time and gravity has attracted much attention 
in recent years\cite{Carlip:2012wa}.

A class of models with such an idea is given by the matrix models. Feynman diagrams generated from
a matrix model can be interpreted as dual diagrams of two-dimensional simplicial manifolds. 
From various aspects, it is established that the matrix models correctly describe the two-dimensional 
gravity\cite{DiFrancesco:1993nw}.
 
A natural extension of the matrix models to higher dimensions was proposed as tensor 
models\cite{Ambjorn:1990ge,Sasakura:1990fs,Godfrey:1990dt}.
They have tensors in place of matrices as their degrees of freedom.
The tensor models generate Feynman diagrams which may be regarded 
as dual diagrams of higher dimensional simplicial manifolds.
However, the original proposal with symmetric tensors has turned out to have various difficulties. 
Recently, a more successful class of tensor models with unsymmetric tensors, called colored tensor models,  
has been proposed and triggered a number of successive studies\cite{Gurau:2011xp}.

In the above description, however, there exists a fundamental issue in regarding the tensor models as theory of emergent space-time
and gravity; the ranks of tensors depend on dimensions of simplicial manifolds. This is serious
because dimensions should be a dynamical quantity rather than an input parameter 
in the idea of emergent space-time.  

A way to circumvent this issue is to interpret rank-three tensor models in another way
than above. 
A fuzzy space 
defines a space in terms of an algebra of functions $f_a\ (a=1,2,\ldots,N)$ on it as
$
f_a f_b=C_{ab}{}^c f_c,
$
where $C_{ab}{}^c$ is the structure constant.
By identifying $C_{ab}{}^c$ with a dynamical variable
of a rank-three tensor model, it may be regarded as a theory
of dynamical fuzzy spaces\cite{Sasakura:2011ma}. 
In this interpretation, a rank-three tensor model
can in principle deal with any dimensional spaces, since fuzzy spaces can approximate 
any continuous spaces. 
This interpretation will be justified, only when its dynamics is shown to 
reproduce a notion of space-time 
with properties required by the nature, 
such as classicality, smoothness, three-dimensionality, locality, causality, symmetries, containing gravity 
and matters. 
This is a long-range purpose of the study of rank-three tensor models.

The main topic here is how to introduce time in a rank-three tensor model by constructing local Hamiltonians. 
A non-trivial issue in the introduction is the mutual consistency of local time evolutions. 
As in the general relativity, this will be guaranteed by a first class constraint algebra formed by
local Hamiltonians and a kinematical symmetry of a rank-three tensor model.

\section{The rank-three tensor model}
The rank-three tensor model in this paper has a rank-three tensor which satisfies the generalized
Hermiticity condition,
\begin{equation}
\label{eq:genHer}
M_{abc}=M_{bca}=M_{cab}=M^*_{bac}=M^*_{acb}=M^*_{cba},
\end{equation}
where $a,b,c=1,2,\cdots,N$.
Because of the generalized Hermiticity condition, the symmetry of the tensor model is the orthogonal
group,
\begin{equation}
\label{eq:orthosym}
M'_{abc}=L_{a}{}^{a'}L_{b}{}^{b'}L_{c}{}^{c'} M_{a'b'c'}, \ \ \ L\in O(N).
\end{equation}

I also introduce the canonical conjugate to $M_{abc}$, which satisfies the Poisson bracket,
$\{ M_{abc},P_{def}\}=\delta_{abc,def}$,
where
\begin{align}
&\delta_{abc,def}\equiv \delta_{ad}\delta_{be}\delta_{cf}+\delta_{ae}\delta_{bf}\delta_{cd}+\delta_{af}\delta_{bd}\delta_{ce},\\
&P_{abc}=P_{bca}=P_{cab}=P^*_{bac}=P^*_{acb}=P^*_{cba}.
\end{align}

The generators of the symmetry (\ref{eq:orthosym}) are given by
\begin{equation}
\label{eq:generator}
{\cal D}_{[ab]} =\frac{1}{2}\left( P_{acd}M_{bcd}-M_{acd}P_{bcd} \right),
\end{equation}
where $[ab]$ denotes anti-symmetric indices.

\section{The interpretation in terms of fuzzy spaces}
A fuzzy space is defined by an algebra of functions $f_a\ (a=1,2,\ldots,N)$ on it,
\begin{equation}
\label{eq:fuzzyalg}
f_a f_b=M_{abc} f_c.
\end{equation} 
Because of the generalized Hermiticity condition (\ref{eq:genHer}), a fuzzy space associated to a
rank-three tensor model is equipped with an inner product which has a trace-like property as
\begin{align}
\langle f_a | f_b \rangle&=\delta_{ab}, \\
\langle f_a|f_b f_c \rangle&=\langle f_a f_b | f_c \rangle=\langle f_c f_a | f_b \rangle,\\
f_a^*&=f_a,\\
(f_af_b)^*&=f_b f_a.
\end{align} 
Note that associativity is not assumed in general. A non-associative 
algebra with such a trace-like inner product may be 
called a non-associative Frobenius algebra. An algebra without associativity does not often tend to
show physically interesting general properties. On the other hand, 
this trace-like feature of the inner product plays key roles in
showing that the kind of fuzzy spaces have various quantum mechanical properties and 
that the symmetry transformation (\ref{eq:orthosym}) can be expressed by $n$-ary 
transformations\cite{Sasakura:2011ma}.
 
I will give here two simple examples of the above kind of fuzzy spaces. 
The function algebra of a usual space is given by
\begin{equation}
\label{eq:usualspace}
f_{z_1}f_{z_2}=\delta^D(z_1-z_2)f_{z_1},
\end{equation}
where $z_i$ are $D$-dimensional coordinates. This can be understood by considering functions
of $x$ which are labeled by $z$, $f_z(x)\equiv\delta^D(x-z)$ and that $f_{z_1}(x)f_{z_2}(x)=
\delta^D(x-z_1)\delta^D(x-z_2)=\delta^D(z_1-z_2)f_{z_1}(x)$. In fact, all the functions of $x$
can be expressed as a linear combination of $f_z$.
Thus, with $M_{abc}$, a usual space can be represented by $M_{z_1z_2z_3}=\delta^D(z_1-z_2)\delta^D(z_2-z_3)$. 
One may consider a fuzzy space by regularizing the delta functions by Gaussians as
\begin{equation}
\label{eq:gauss}
M_{z_1z_2z_3}=B \exp\left(-\beta\left((z_1-z_2)^2+(z_2-z_3)^2+(z_3-z_1)^2\right)\right),
\end{equation}
where $B,\beta$ are positive numbers.  Since this respects the $D$-dimensional Poincar\'e symmetry,
the fuzzy space should be regarded as a fuzzy $D$-dimensional flat space.

\section{Symmetry of tensor models and diffeomorphism on fuzzy spaces}
From the relation (\ref{eq:fuzzyalg}), the symmetry transformation (\ref{eq:orthosym}) of the tensor model 
is mapped to a transformation of a fuzzy space, 
\begin{equation}
\label{eq:onfuzzy}
f_a'=L_{a}{}^{a'}f_{a'},\ \ L\in O(N).
\end{equation} 
Since the functions $f_a$ can intuitively be considered to represent fuzzy ``points",
the transformation (\ref{eq:onfuzzy}) can be
regarded as a fuzzy analogue of coordinate transformation.

One can discuss this more precisely. The generators (\ref{eq:generator}) satisfy the $so(N)$ Lie algebra,
\begin{align}
\label{eq:ddd}
&\{ D(V_1),D(V_2)\}=D([V_1,V_2]),\\
&D(V)\equiv V^{[ab]}{\cal D}_{[ab]},
\end{align}
where $V^{[ab]}$ denotes an anti-symmetric matrix, and $[V_1,V_2]$ is the matrix commutator. 
Now let me formally replace the indices with
$D$-dimensional coordinates, and consider
\begin{equation}
V^{[xy]}\equiv \frac{1}{2} \left( v^\mu(x)+v^\mu(y)\right)\delta^D_\mu(x-y),
\end{equation}
which is parameterized by a vector field $v^\mu(x)$. Then one can find that $V_3=[V_1,V_2]$ leads to
\begin{equation}
v_3^\mu(x)=v_1^\nu(x) \partial_\nu v_2^\mu(x)-v_2^\nu(x) \partial_\nu v_1^\mu(x)={\cal L}_{v_1}v_2.
\end{equation} 
This is indeed the algebra of diffeomorphism.

\section{Construction of local Hamiltonians}
Mutual consistency of local time evolutions
requires that commutation of local Hamiltonians should vanish up to symmetries, namely,  
\begin{equation}
\label{eq:genrequire}
\{ {\cal H}_a,{\cal H}_b \} =\hbox{symmetry generators}.
\end{equation}
In the case of tensor models, there actually exists a large symmetry (\ref{eq:orthosym}), where $N$ is 
intuitively the number of ``points".

The search for such consistent local Hamiltonians is a complicated process. The final result 
\cite{Sasakura:2012fb} is that
there exist only two such Hamiltonians,
\begin{equation}
\label{eq:localham}
{\cal H}_a=P_{a(bc)}P_{bde}M_{cde}\ \hbox{or}\  {\cal H}_a=P_{a(bc)}P_{bde}M_{ced},
\end{equation}
with $(ab)$ denoting symmetrization of the indices,
which have a slight difference in index contraction,
on the following assumptions: (A1) Local Hamiltonian has one index. (A2) Form a consistent
first class constraint algebra. (A3) Invariant under time reversal. (A4) At most cubic polynomial 
function. (A5) Terms are connected.  

\section{First class constraint algebra and DeWitt algebra}
The local Hamiltonians (\ref{eq:localham}) and the $so(N)$ generators (\ref{eq:generator}) satisfy
\begin{align}
\label{eq:halg}
&\{ H(T_1), H(T_2)\}=D([\tilde T_1,\tilde T_2]), \\
&\{ D(V), H(T)\}=H(VT),
\end{align}
where $H(T)\equiv T^a {\cal H}_a,\ \tilde T_{ab}\equiv T^cP_{c(ab)}$.  
With (\ref{eq:ddd}), they form a first class constraint algebra. 
Note that (\ref{eq:halg}) has structure functions rather than constants, because of  
the dependence of $\tilde T$ on $P_{abc}$. This means that the algebra is in fact an open algebra, which
closes only on the constraint surfaces.

This situation is very similar to the general relativity. In fact, one can show that, by formally replacing
the indices with coordinates, substituting (\ref{eq:gauss}) into $P_{abc}$ in the 
right-hand side of (\ref{eq:halg}), and taking $\beta\rightarrow\infty$ limit, one reproduces the DeWitt constraint algebra of 
general relativity \cite{Sasakura:2011sq}.

\section{Quantization}
A way of quantization is to promote the canonical variables to the operators satisfying
$[\hat M_{abc},\hat P_{def}]=i \delta_{abc,def}$.
Then the local Hamiltonians are generally corrected by operator ordering as
\begin{equation}
\hat {\cal H}_a=\hat P_{a(bc)} \hat P_{bde}\hat M_{cde} + i \lambda \hat P_{abb}.
\end{equation}
This is similar to the other Hamiltonian. One can see that the first class constraint algebra
 is not deformed by the quantization. 
 Therefore the algebra gives a consistent quantization of the system. 

Another way of quantization is to apply the path integral quantization of constraint systems \cite{Faddeev:1969su}.  It can formally be written as 
\begin{equation}
Z=\int {\cal D}M {\cal D}P\, \delta({\cal H}_a) \delta({\cal D}_{ab}) \delta(\chi_a)
\delta(\chi_{ab}) \hbox{Det}(\{ {\cal H}\, {\cal D},\chi \})e^{i\int PdM},
\end{equation}
where $\chi_a,\chi_{ab}$ are gauge-fixing conditions.

\section{Summary and future prospects} 
A canonical tensor model with the notion of local time has been proposed. The consistency of local time evolutions
is guaranteed by a first class constraint algebra among local Hamiltonians and symmetry generators.
The two-fold uniqueness of local Hamiltonians has been proved on some physically reasonable assumptions.
It would be highly interesting to study whether the quantization of the canonical tensor model
leads to the emergence of space-time with the features required by the nature. 

It would also be highly interesting to seek for connections to the CDT\cite{Ambjorn:2012jv}.
 
\section*{Acknowledgments}
I would like to thank the organizers for this nice opportunity to present this work at the workshop.

\end{document}